\documentclass[prl,twocolumn,amsmath]{revtex4}

\usepackage{graphicx}
\usepackage{amssymb,amsfonts,amsmath}

\def \>{\rangle} 
\def \<{\langle} 
 
\def\be{\begin{equation}} 
\def\ee{\end{equation}} 
\def\longrightharpoonup{\relbar\joinrel\rightharpoonup}
\def\longleftharpoondown{\leftharpoondown\joinrel\relbar}

\def\longrightleftharpoons{
  \mathop{
    \vcenter{
      \hbox{
      \ooalign{
        \raise1pt\hbox{$\longrightharpoonup\joinrel$}\crcr
	  \lower1pt\hbox{$\longleftharpoondown\joinrel$}
	  }
      }
    }
  }
}

\newcommand \bea {\begin{eqnarray}} 
\newcommand \eea {\end{eqnarray}}

\newcommand{\bs}{\boldsymbol{\sigma}}

\begin{document}

\title{ Zipf's law and criticality in multivariate data without
  fine-tuning }

\author{David J. Schwab}
\email{dschwab@princeton.edu}
\affiliation{Department of Physics and Lewis-Sigler Institute,
  Princeton University, Princeton, NJ 08854} 
\author{Ilya Nemenman}
\email{ilya.nemenman@emory.edu}
\affiliation{Departments of Physics and Biology, Emory University,
  Atlanta, GA 30322} 
\author{Pankaj Mehta}
\email{pankajm@bu.edu}
\affiliation{Department of Physics, Boston University, Boston, MA
  02215}

\begin{abstract}
The joint probability distribution of many degrees of freedom in biological systems, such as firing patterns in neural networks or antibody sequence composition in zebrafish, often follow Zipf's law, where a power law is observed on a rank-frequency plot. This behavior has recently been shown to imply that these systems reside near to a unique critical point where the extensive parts of the entropy and energy are exactly equal. Here we show analytically, and via numerical simulations, that Zipf-like probability distributions arise naturally if there is an unobserved variable (or variables) that affects the system, e. g. for neural networks an input stimulus that causes individual neurons in the network to fire at time-varying rates. In statistics and machine learning, these models are called latent-variable or mixture models. Our model shows that no fine-tuning is required, i.e. Zipf's law arises generically without tuning parameters to a point, and gives insight into the ubiquity of Zipf's law in a  wide range of systems.
  %It does not require fine tuning, but suggests than biological systems must be adapted to the extrinsic stimulus statistics to exhibit the Zipf's law. 
\end{abstract}

\maketitle

Advances in high throughput experimental biology now allow the
joint measurement of activities of many basic components
underlying collective behaviors in biological systems. These include
firing patterns of many neurons responding to a movie
\cite{Gaspar,cocco2009neuronal, schneidman2006weak, tkavcik2013simplest},
sequences of proteins from individual immune cells in zebrafish
\cite{mora2010maximum, murugan2012statistical}, protein sequences more generally
\cite{weigt2009identification, halabi2009protein}, and even the
simultaneous motion of flocking birds \cite{bialek2012statistical}. A
remarkable result of these data and their models
has been the observation that these large biological systems often reside close to a critical point \cite{Gaspar,mora2011biological}.
This is most clearly manifest directly from the data by the
following striking behavior. If we order the states, $\bs$, of a system
by decreasing probability, then the frequency of the states decays as
the inverse of their rank, $r(\boldsymbol{\sigma})$, to some power:
\begin{equation}
  P(\boldsymbol{\sigma}) \propto
  \frac{1}{r(\boldsymbol{\sigma})^\alpha}.
\label{Zipflaw}
\end{equation}
Many systems in fact exhibit
$\alpha\simeq1$, which is termed Zipf's law, and on which we will
focus.

It has been argued that Zipf's law is a
model-free signature of criticality in the underlying system, using the language of statistical mechanics
\cite{mora2011biological}. Without loss of generality, we can define
the ``energy'' of a state $\bs$ to be
\begin{equation}
E(\boldsymbol{\sigma}) =
-\log{P(\boldsymbol{\sigma})} + \mbox{const.}
\label{defEnergy}
\end{equation}
The additive constant is arbitrary, and the temperature is $k_B
T=1$. We can also define the ``entropy'', $S(E)$, using the density of
states, $\rho(E) =\sum_{\boldsymbol{\sigma}}
\delta(E-E(\boldsymbol{\sigma}))$, as \be S(E)= \log{\rho(E)}.
\label{defEntropy}
\ee Both the energy $E$ and the entropy $S(E)$ contain extensive terms
that scale with the system size, $N$. An elegant argument
\cite{mora2011biological} converts Eq.~\!(\ref{Zipflaw}) with $\alpha=1$
into the statement that, for a large system, $N \to\infty$, the energy
and entropy are exactly equal (up to a constant) to leading order in $N$. Thus in the
thermodynamic limit, the probability distribution is indeed poised
near a critical point where all derivatives beyond
the first of the entropy with respect to energy vanish to leading
order in $N$.

The observation of Zipf's law in myriad distributions inferred from
biological data has contributed to a revival of the idea that
biological systems may be poised near a phase transition
\cite{mora2011biological,bak,beggs,beggs2,Kitzbichler,Chialvo}. Yet most
existing mechanisms to generate Zipf's law can produce a variety of
power-law exponents $\alpha$ (see \cite{newman2005power,
  clauset2009power} and reference therein), have semi-stringent
conditions \cite{dutta}, are domain-specific, or require fine-tuning to a critical point,
highlighting the crucial need to understand how Zipf's law can
arise in data-driven models.

Here we present a generic mechanism that produces Zipf's law and does not
require fine-tuning. The observation motivating this new mechanism is
that the correlations measured in biological data sets have multiple
origins. Some of these are intrinsic to the system, while the others
reflect extrinsic, unobserved sources of variation \cite{Marsili,Munoz}. For example, the
distributions of activities recorded from networks of neurons in the
retina reflect both the intrinsic structure of the network as well as
the stimuli the neurons receive \cite{Macke}, such as a movie of natural
scenes. Likewise, in the immune system, the pathogen environment is an
external source of variation that influences the observed antibody
combinations. We will show that the presence of such unobserved,
hidden random variables naturally leads to Zipf's law. Unlike other
mechanisms \cite{newman2005power,dutta}, our approach requires a large
parameter (i.\ e., the system size, or the number of observations), with power-law behavior emerging only in the thermodynamic
limit. On the other hand, our mechanism does not require fine-tuning of parameters to a point
or any special statistics of the hidden variables \cite{Hertz}. In
other words, Zipf's law is a universal feature that emerges when
marginalizing over relevant hidden variables.

{\em A simple model --- } In order to understand how a hidden variable
can give rise to Zipf's law and concomitant criticality, we start by
examining a simple case of $N$ conditionally independent binary spins
$\sigma_i=\pm 1$. The spins are influenced by a hidden variable
$h$ drawn from a probability distribution $q(h)$, which is
smooth and independent of $N$. In particular, we consider the case
\begin{equation}
P(\boldsymbol{\sigma}|h) = \prod_{i=1}^N P(\sigma_i | h) =  \prod_{i=1}^N \frac{e^{h \sigma_i}}{2 \cosh{h}}.
\label{identicalspins}
\end{equation}
Note that our chosen form of $P(\sigma_i | h)$ imposes no loss of generality for non-interacting binary variables. We consider a scenario where the parameter $h$ changes rapidly
compared to the duration of the experiment, so that the probability distribution of
the measured data, $\bs$, is averaged over $h$:
\begin{align}
P(\boldsymbol{\sigma})&= \frac{1}{2^N} \int dh\, q(h) e^{N
  (h m(\boldsymbol{\sigma})- \log{\cosh{h}})}\\&\equiv  \frac{1}{2^N} \int dh\,
q(h) e^{-N{\mathcal H}(m,h)},
\label{MarginalProb}
\end{align}
where we have defined the average magnetization $m= \sum_i \sigma_i
/N$, and the last equation defines
${\mathcal H}(m,h)$. Note that the distribution $P(\bs)$ does not
factorize unlike $P(\bs|h)$. That is, the conditionally
independent spins are not marginally independent. Indeed, as in
\cite{bnt}, a sequence of spins carries information about the
underlying $h$ and hence about other spins (e.~g., a prevalence of
positive spins suggests $h>0$, and thus subsequent spins will also
likely be positive). We note that the simple model in
Eq.~\!(\ref{MarginalProb}) is intimately related to the MaxEnt model
constructed in \cite{tkavcik2013simplest} to match the distribution of
the number of simultaneously firing retinal ganglion cells.

In the limit $N \gg 1$, we can approximate the integral in
Eq.~\!(\ref{MarginalProb}) by Laplace's method (saddle-point
approximation) to get
\begin{equation}
P(\boldsymbol{\sigma}) \approx {2^{-N}} q(h^*) e^{N (h^* m- \log{\cosh{h^*}})},\;
\tanh{h^*} =m.
\label{saddle}
\end{equation}
Here $h^*$ is the maximum-likelihood estimate of $h$ given the
data, $\boldsymbol{\sigma}$. In deriving Eq.~\!(\ref{saddle}) we assumed
that the distribution $q(h)$ has support at $h^*$ and is
sufficiently smooth, e.g. does not depend on $N$, so that the
saddle-point over $h$ is determined by ${\mathcal H}$, and not by
the prior.  In other words, we require the Fisher information
${\mathcal F}({h^*})\equiv -N\left.\frac{\partial^2 {\mathcal
      H}}{\partial h^2}\right|_{h^*}=N(1-m^2)\gg1$, and for the
location and curvature of the saddle point to not be
significantly modulated by $q(h$).  These conditions are violated
at $m = \pm 1$, and there is a semi-infinite range of $h$ that
could have contributed to such states. For all nonzero values of
${\mathcal F}$, the saddle-point will eventually dominate over
$q(h)$ as $N\to\infty$. However, the convergence is not uniform.

Substituting Eq.~\!(\ref{saddle}) into Eq.~\!(\ref{defEnergy}) and using
the identities $\tanh^{-1} m = \frac{1}{2}\log{\left({1+m \over
      1-m}\right)}$ and $\cosh{[\tanh^{-1}m]}=(1-m^2)^{-1/2}$, we
obtain the
energy to leading order in $N$:
\begin{align}
E(m)&\approx\textstyle
  -N\left[\left(\frac{1+m}{2}\right)\log{\left(\frac{1+m}{2}\right)}
    +\left(\frac{1-m}{2}\right)\log{\left(\frac{1-m}{2}\right)}
  \right]\nonumber\\&\equiv NH(m). \label{EnergyIndependent}
\end{align}
Here we neglected subdominant terms that come from both the prior
$q(h^*)$ and the fluctuations about the saddle point.  It is worth
noting that this energy considered as a function of the $\sigma_i$,
rather than $m$, includes interactions of all orders, not just
pairwise spin couplings.

We can also calculate the entropy $S(m)$ associated with the
magnetization $m$. For a system of $N$ binary spins, each state with
magnetization $m$ has $K=N\left(\frac{1+m}{2}\right)$ up spins, and
there are ${N \choose K}$ such states. Using Stirling's
approximation, one finds that the entropy takes the familiar form $
S(m)=\log{{N \choose K}} \approx NH(m)$. Of course, this is the same
as the energy, Eq.~\!(\ref{EnergyIndependent}), for the system with a
hidden variable $\beta$, to leading order in $N$.

\begin{figure}[t]
\centering
\includegraphics[width=0.75\columnwidth]{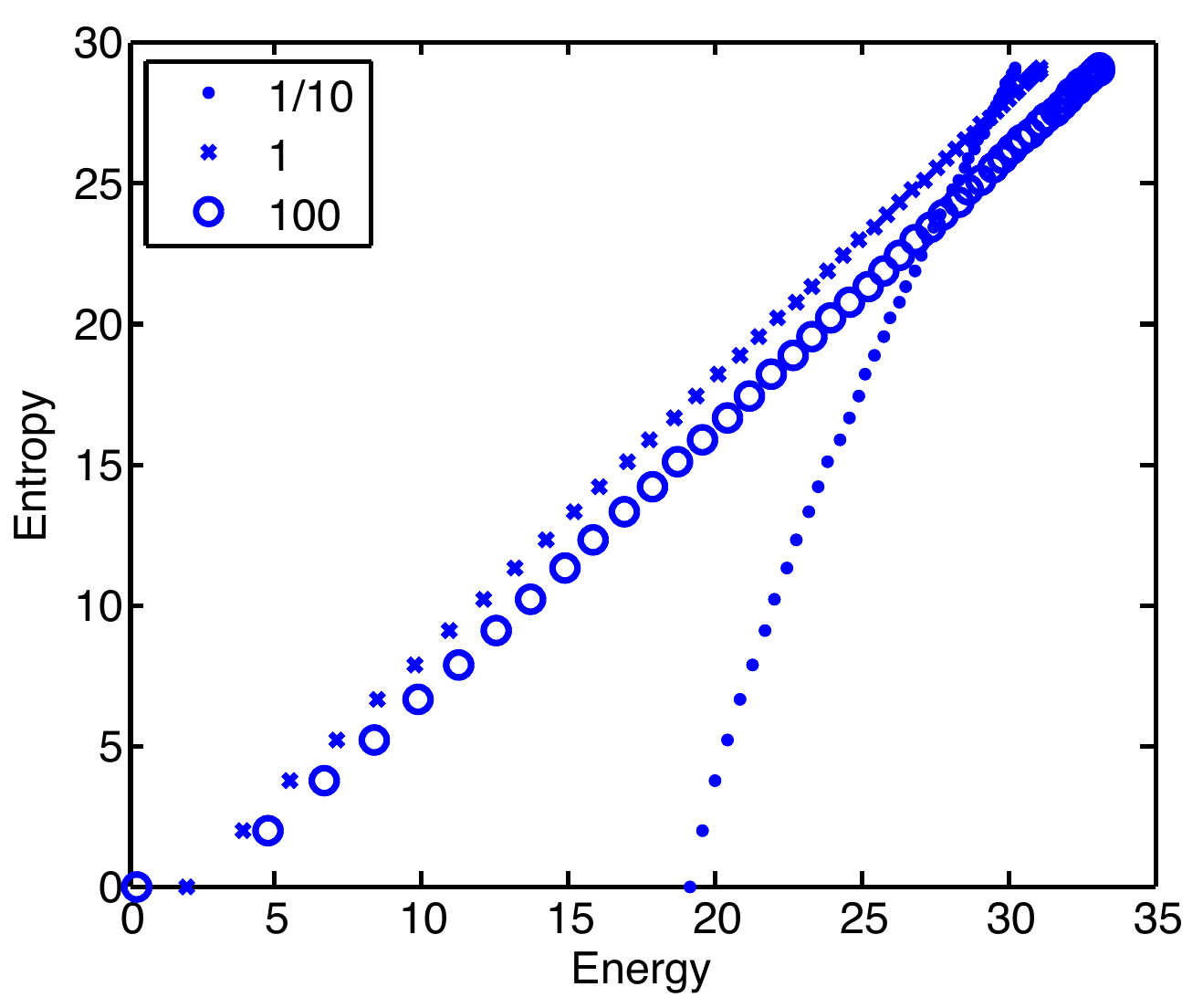}
\caption{\label{EvsS} Entropy, $S(m)$, vs energy, $E(m)$, for $N=100$
  identical and conditionally independent spins. Zipf's law ($E=S$)
  emerges as the standard deviation, $s \in \{0.1,1,100\}$, of the
  Gaussian distribution characterizing the hidden variable $h$ is
  increased. Notice that there is a nearly perfect Zipf's law for $2$
  orders of magnitude in $s$. The mean of $q(h)$ is set to zero,
  and thus there is a two-fold degeneracy between states with
  magnetization $m$ and $-m$. }
\end{figure}

The analytic equivalence between energy and entropy only applies when
$N \rightarrow \infty$. To verify our result for a finite $N$, we
numerically calculate $E(m)$ from Eq.~\!(\ref{MarginalProb}) with
$q(\beta)$ chosen from a variety of distribution families (e.~g.,
Gaussian, exponential, uniform). For brevity, we only show plots for
Gaussian distributions, but the others gave similar results.  Figure
\ref{EvsS} plots the entropy, $S(m)=\log{{N \choose K}} $, vs the
energy, $E(m)$, for $N=100$ conditionally independent spins, where
$q(h)$ has mean $0$ and varying standard deviation $s \in
\{0.1,1,100\}$. For small $s$, the hidden variable $h$ is always
close to zero, there is no averaging, and all states are nearly
equally (im)\-probable.  As $s$ increases, entropy becomes equal to
energy over many decades of energies modulo an arbitrary additive
constant. This holds true for two orders of magnitude of the standard
deviation $s$, confirming that our mechanism does not require fine
tuning.
% It should be noted that while we have chosen to call the hidden
% variable $\beta$ in analogy to inverse temperature, here we allow it
% to take negative values. In principle, one can also construct a
% frequency vs. rank log-log plot, i.e. a ``Zipf plot'', for each
% configuration of $\boldsymbol{\sigma}$. One would find a series of
% degenerate plateaus, due to the identical nature of the spins, that
% nonetheless follow Zipf's law at the level of the steps. We will
% show below that such degeneracy is not necessary for our mechanism,
% and that the addition of quenched heterogeneity smooths these steps.

The stable emergence
in the thermodynamic limit, $N\to\infty$, with no fine-tuning,
distinguishes our setup from a classic mechanism explaining $1/f$
noise in solids \cite{dutta} and certain other biological systems
\cite{tu}. We could have anticipated this result: if the extensive
parts of the energy and entropy do not cancel, in thermodynamic limit,
the magnetization will be sharply peaked around the $m$ that
minimizes the free-energy, $Nf(m) = E(m)-S(m)$. Thus in order for there to be a broad
distribution of magnetizations within $P(\bs)$ the extensive part of
$f(m)$ must be a constant. In other words, the
observation of a broad distribution of an order parameter-like
quantity in data is indicative of a Zipfian distribution. One
straightforward mechanism to produce a broad order parameter
distribution for large $N$ is to couple it to a hidden fluctuating
variable.

{\em A generic model --- }We now
show that Zipf-like criticality is a generic property of
distributions with hidden variables, and is not a consequence of the
specific model in Eq.~\!(\ref{identicalspins}). In particular, it does
not require the observed variables to be identical or
conditionally independent, nor the fluctuating parameter(s) to be temperature-like.

Consider a probabilistic model of data, ${\bf x} = (x_1,x_2,...,x_N)$, with $M$ parameters, ${\bf g}=(g_1,\ldots ,g_M)$. Without loss of generality, \footnote{Any distribution can be written in this form. Moreover, our derivation does not make use of the structure of the parameter distribution, $Q({\bf g})$. In particular the parameters are not required to be independent.} we can write the probability distribution in the log-linear form
\begin{equation}
P(x_1,x_2,...,x_N|{\bf g}) =\frac{1}{Z({\bf g})}\exp\left[-N\sum_{\mu=1}^{M}g_\mu \mathcal{O}_\mu({\bf x})\right],
\end{equation}
where we have defined the partition function 
\begin{equation}
Z({\bf g}) = \int d^N x' \exp\left[-N\sum_{\mu=1}^{M}g_\mu \mathcal{O}_\mu({\bf x'})\right].
\end{equation}
If the $x_i$ are discrete, the integral is instead a summation. As an example, the fully-connected Ising model would have ${\bf g} = (h,J)$, with $\mathcal{O}_1=-\frac{1}{N}\sum_i x_i$ and $\mathcal{O}_2=-\frac{1}{N^2}\sum_{i<j}x_i x_j$, with each $x_i \in \{-1,1\}$.

If  the first $K$ out of the $M$ parameters are chosen to fluctuate, according to a distribution $Q({\bf g})$, then the marginal distribution of the data, ${\bf x}$, is given by
\begin{equation}
P({\bf x}) = \int d^K g \, Q(g_1,g_2,...,g_K)e^{-N F({\bf g},{\bf x})},
\label{marginal}
\end{equation}
with
$F({\bf g};{\bf x}) = \sum_{\mu=1}^{M}g_\mu \mathcal{O}_\mu({\bf x})+\frac{1}{N} \ln Z({\bf g})$.
If the distribution of the $K$ fluctuating variables, $Q({\bf g})$, is sufficiently broad, as discussed after Eq.~\!(\ref{saddle}), we can perform a saddle-point approximation to this integral. Denote the solution to the saddle-point equations by ${\bf g}^*=(g_1^*,...,g_K^*,g_{K+1},...,g_M)$ \footnote{If we again consider the fully-connected Ising model,  choosing $K=1$ with $g_1=h$, $g_2=J$, then there exists a solution to the saddle-point equations for any $J\leq1$. For $J>1$, however, there is a first-order phase transition at $h=0$, with a jump in the magnetization, and sufficiently small values of the average magnetization cannot be accessed by tuning $h$.}. Neglecting subleading terms, the saddle-point approximation to the integral yields
\begin{equation}
E({\bf x}):= -\frac{1}{N} \ln P({\bf x}) = \sum_\mu g_\mu^* \mathcal{O}_\mu({\bf x}) + \frac{1}{N}\ln Z ({\bf g}^*),
\label{generalE}
\end{equation}
where ${\bf g}^*$ is the solution to
\begin{equation}
\frac{1}{N}\frac{\partial  \ln Z({\bf g})}{\partial g_\nu}|_{\bf g^*} = -\mathcal{O}_\nu({\bf x})
\end{equation}
for $\nu=1...K$.
Notice that the  $g_\mu^*$ are functions of the data through the $\mathcal{O}_\mu({\bf x})$.

We can compare the energy in Eq.~\!(\ref{generalE}) to the microcanonical entropy, $S(\{\mathcal{O}_\mu({\bf x})\})$, calculated empirically from the data. For our problem, the multi-dimensional form of the G\"{a}rtner-Ellis theorem \cite{van1991collective} states that the entropy
\begin{equation}
S(\{\mathcal{O}_\mu({\bf x})\}) = \inf_g \left[\sum_\mu g_\mu \mathcal{O}_\mu({\bf x})+c({\bf g})\right]
\label{fullS}
\end{equation}
is the Legendre-Fenchel transform of the cumulant generating function, which is, aside from an unimportant constant, just minus the free-energy,
\begin{equation}
c({\bf g}) = \lim_{N\to \infty} N^{-1}\ln Z({\bf g}) - C,
\end{equation}
where $C= \frac{1}{N}\ln \int d^N x'$. If $K=M$, Eq.~\!(\ref{fullS}) is identical to Eq.~\!(\ref{generalE}), and we have proven Zipf's law, i.e. $S(\{\mathcal{O}_\mu({\bf x})\}) = E({\bf x})$.

Even if $K<M$, if satisfying Eq.~\!(13) for $\nu=1...K$  automatically satisfies Eq.~\!(13) for $\nu=K+1...M$, then Zipf's law will hold. For example, in the fully-connected Ising model, matching the average magnetization, $m$, automatically provides matching of the pairwise interaction term, since it's simply $m^2/2$. In other words, Zipf's law will hold for any form of static interactions if the expectation values of operators conjugate to the static parameters are functions of the expectation values of operators conjugate to the fluctuating parameters.

\begin{figure}[t]
\centering
\includegraphics[width=0.8\columnwidth]{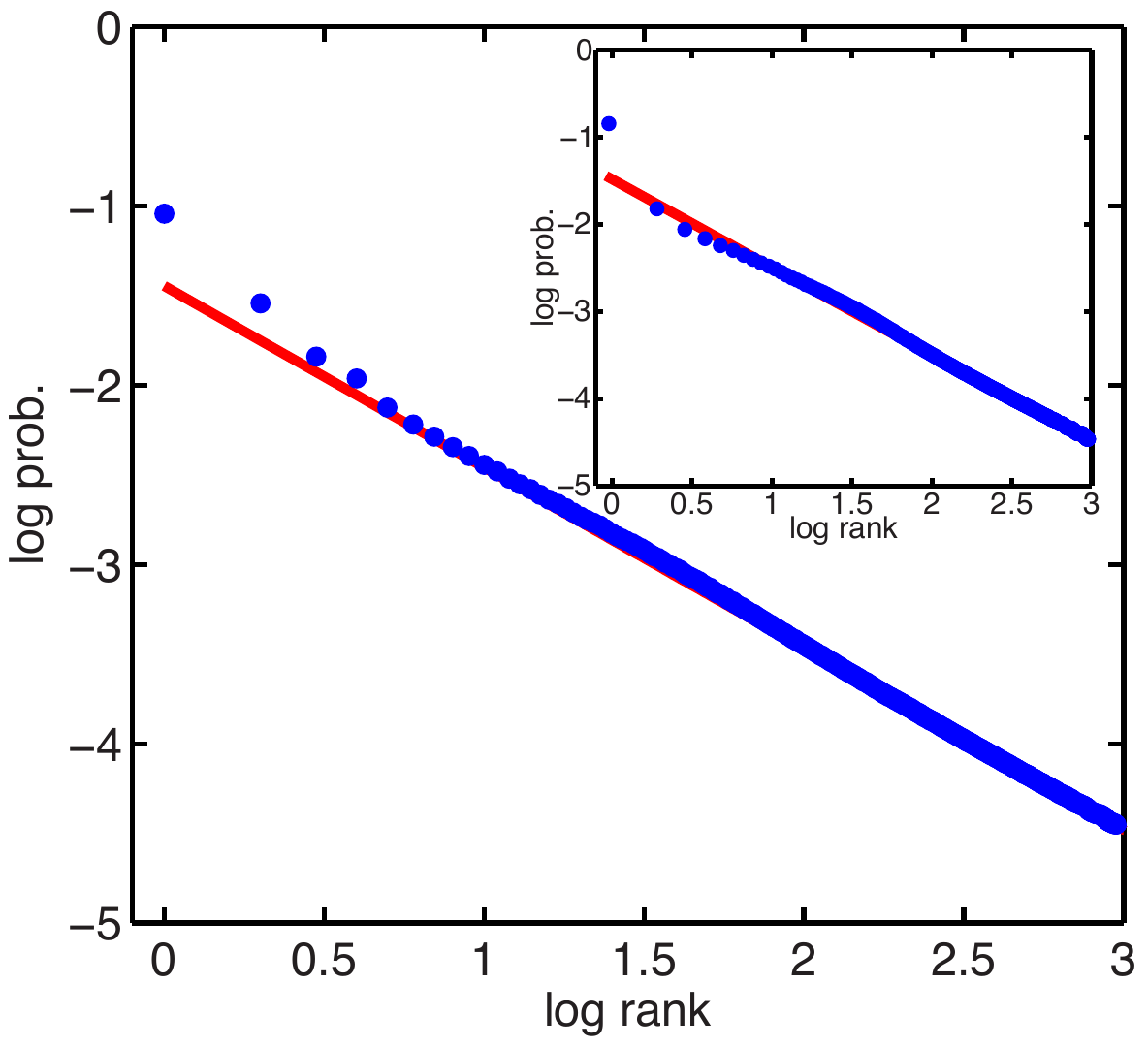}
\caption{ \label{zipf} Main plot: Plot of $\log_{10}$ probability
  vs. $\log_{10}$ rank of the most frequent $10^3$ states for a system
  of $N=200$ non-identical, conditionally-independent spins (model
  (a)). Plots are an average over 200 realizations of the quenched
  variables $h_i$ that break the symmetry between spins, with
  $5\times10^5$ samples taken for each realization. Parameters: $h_i
  \sim {\cal N}(\mu=1, s=0.3)$, $\beta \sim {\cal N}(\mu=0, s=2)$. Red
  line: least-squares fit to patterns $100-1000$, slope of
  $-1.012$. Inset: Same as above, except for a model of $N=200$ spins
  with quenched random interactions $J_{ij}$ and biases $h_i$ (model
  (b)). Average over $10$ realizations of $J_{ij}$ and $h_i$ chosen
  from $J_{ij} \sim {\cal N}(\mu=1, s=0.5)$, $h_i \sim {\cal
    N}(\mu=1, s=0.85)$, $\beta \sim {\cal N}(\mu=0.5,
  s=0.5)$, with $3\times10^5$ samples taken for each
  realization. Red line: least-squares fit to patterns $100-1000$,
  slope of $-1.011$.}
\end{figure}

We numerically test the validity of our analytic result for finite $N$
in two systems more complex than Eq.~\!(\ref{identicalspins}): (a) a
collection of non-identical but conditionally independent spins, and
(b) an Ising model with random interactions and fields. The main graph
of Fig.~\ref{zipf} shows a Zipf plot for system (a), so that \be
P(\boldsymbol{\sigma}|\beta) = \prod_{i=1}^N P(\sigma_i | \beta) =
\prod_{i=1}^N \frac{e^{-\beta h_i \sigma_i}}{2 \cosh{\beta h_i}},
\label{identicalspins2}
\ee where $h_i$ are quenched, Gaussian random variables
unique for each spin. In the simulations, the hidden variable $\beta$
was drawn from a Gaussian distribution, but similar results were found
for other distributions. The quenched fields $h_i$ break the symmetry
between spins. In agreement with our derivation, on a log-log
plot, the states generated from simulations fall on a line with slope
very close to $-1$ (Fig.~\ref{zipf}), the signature of Zipf's
law.

To verify that conditional independence is not required for this
mechanism, we studied system (b) that generalizes the model in
Eq.~\!(\ref{identicalspins2}) to include random exchange interactions
between spins:
\begin{equation}
P(\boldsymbol{\sigma}|\beta) \propto e^{- \beta \left( \frac{1}{N}
    \sum_{i \neq j} J_{ij} \sigma_i \sigma_j + \sum_i h_i \sigma_i
  \right)},% +\sum_{i\neq j}J^{(0)}_{ij}\sigma_i\sigma_j},
\label{glass}
\end{equation}
where the $J_{ij}$ and $h_i$ are quenched Gaussian distributed
interactions and fields, and $\beta$ is as above. As shown in Fig.~2
(inset), the data again fall on a line with slope nearly equal to
$-1$.

\begin{figure}[t]
\centering
\includegraphics[width=0.8\columnwidth]{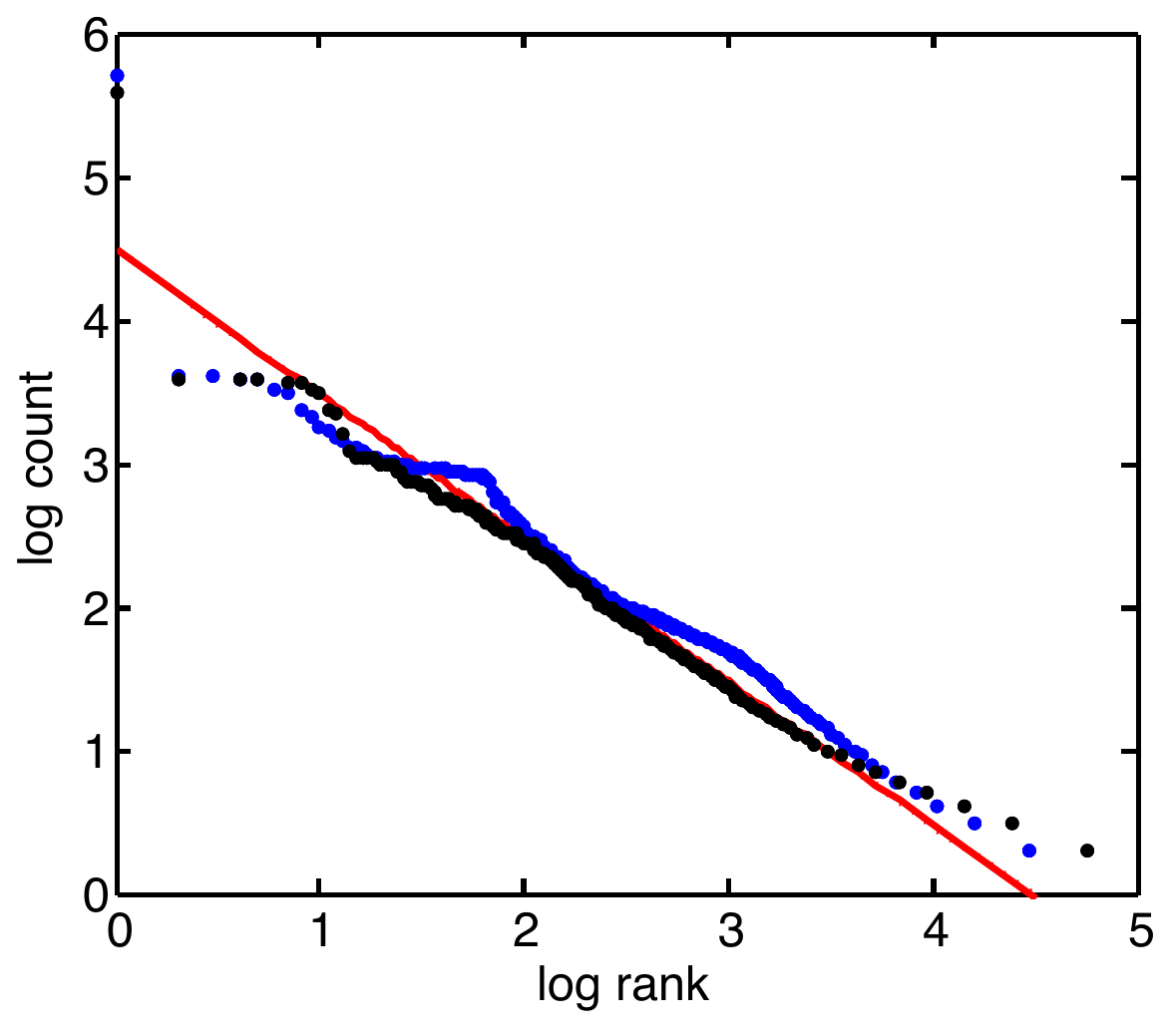}
\caption{ \label{data} Rank-count plot from a motion-sensitive
  blowfly neuron, logs base $10$; discretization is $\tau=1$ ms, and $N=40$.  Black:
  empirical rank-ordered counts. Blue: rank-ordered counts from a simulated refractory Poisson spike train with the input stimulus the same as in the experiment, and with mean firing rate and refractory period matched to the experimental data. Red: slope
  of $-1$ guide to the eye.}
\end{figure}

To see our mechanism at work in data, consider a neural spike train
from a single blowfly motion-sensitive neuron H1 stimulated by a time-varying
motion stimulus, $v(t)$ (see \cite{Nemenman:2004,Nemenman:2008}
for experimental details). We can discretize time with a resolution of
$\tau$ and interpret the spike train as an ordered sequence of $N$
spins, such that $\sigma_i=\pm1$ corresponds to the absence/presence
of a spike in a time window $t\in [\tau(i-1),\tau i)$. The probability
of a spike in a time window depends on $v$. However, neural
refractoriness prevents two spikes from being close to each other,
irrespective of the stimulus, resulting in a repulsion that does not
couple to $v$.  The rank-ordered plot of spike patterns produced
by the neuron is remarkably close to the Zipf behavior
(Fig.~\ref{data}).  We also simulated a refractory Poisson spike train
using the same values of $v(t)$. We chose the probability of spiking (spin up) as in Eq.~\!(4) with $h(t)=av(t)$, $a={\rm const}$, and with a hard repulsive constraint between positive spins extending over a refractory period of duration $\tau_r$. We then choose $\tau_r$ as the shortest empirical interspike interval ($\approx 2$ ms) , and set $a$ such that the magnetization (the mean firing rate) matches the data. The
rank-ordered plot for this model that manifestly includes interactions
uncoupled from the hidden stimulus $v(t)$ still exhibits Zipf's
law (Fig.~\ref{data}).

{\em Discussion --- }%The observation that a broad class of
%multivariate biological systems appear to be tuned near a critical
%point has renewed interest in the mechanisms that can give rise to
%criticality and Zipf's law \cite{mora2011biological,beggs}. 
%While there exist many known mechanisms to generate power laws, most allow
%for a variety of power law exponents \cite{newman2005power,
 % clauset2009power} and must be tuned to give rise to an
%exponent of strictly $-1$. 
%As has been discussed extensively, critical
%systems have a wide variety of desirable information processing
%properties \cite{mandelbrot,carlson,beggs2}. 
It is possible that
evolution has tuned biological systems or exploited natural 
mechanisms of self-organization \cite{bak} to arrive at Zipf's law. Alternatively, informative 
data-driven models may lie close to a critical point due to the high density
of distinguishable models there \cite{mastromatteo,Marsili}. Our work suggests 
another possibility: Zipf's law can robustly emerge due to
the effects of unobserved hidden variables. While our approach is
biologically motivated, it is likely to be relevant to other systems
where Zipf's law has been observed, and it will be interesting to
unearth the dominant mechanisms in particular systems. For
this, if a candidate extrinsic variable can be identified, such as the
input stimulus to a network of neurons, its variance could be
modulated experimentally as in Fig.~1. Our mechanism would expect
Zipf's law to appear only for a broad distribution of the extrinsic
variable, and for $N\gg 1$ observed variables.

While our mechanism does not require fine-tuning, it nonetheless
suggests that biological systems operate in a special regime. Indeed,
the system size $N$ required to exhibit Zipf's law depends on the
sensitivity of the observed $\boldsymbol{\sigma}$ to the variations of
the hidden variable. If the system is poorly adapted to the
distribution of $h$, e.~g. the mean of $q(h)$ is very large or
its width is too small to cause substantial variability in $\bs$ (as
in Fig.~1, $s=0.1$), a very large $N$ will be required. In other
words, a biological system must be sufficiently adapted to the
statistics of $h$ for Zipf's law to be observed at modest system
sizes. Indeed, this type of adaptation is well established in both
neural and molecular systems
\cite{Laughlin:1981,Brenner:2000,Berg:2004,Nemenman:2012}.

\begin{acknowledgments} {\bf Acknowledgments} We would like to thank
  Bill Bialek, Justin Kinney, H.G.E. Hentschel, Thierry Mora, and Martin Tchernookov for
  useful conversations. We thank Robert de Ruyter van Steveninck 
  and Geoff Lewen for providing the data in Fig. \ref{data}. PM was partially
  supported by an Alfred Sloan Fellowship. IN was partially supported
  by the James S.\ McDonnell Foundation. DJS was supported by National
  Institute of Health Grant K25 GM098875-02. DJS and IN thank the Aspen Center for Physics for their hospitality when this work began.\end{acknowledgments}

\bibliography{pankajdavidrefs}   
\end{document}